\documentclass[reprint,twocolumn,showpacs,superscriptaddress,groupedaddress]{revtex4}  
\usepackage{graphicx}  
\usepackage{dcolumn}   
\usepackage{bm}        
\usepackage{amssymb}   
\usepackage{color}   
\usepackage{amsmath}
\usepackage{amsfonts}
\usepackage{upgreek}
\usepackage{ulem}

\newcommand{\beginsupplement}{%
        \setcounter{table}{0}
        \renewcommand{\thetable}{S\arabic{table}}%
        \setcounter{figure}{0}
        \renewcommand{\thefigure}{S\arabic{figure}}%
        \setcounter{equation}{0}
        \renewcommand{\theequation}{S\arabic{equation}}%
     }

\begin{document}

\title{Controlled uniform coating from the interplay of Marangoni flows and surface-adsorbed macromolecules}

\author{Hyoungsoo Kim}
\affiliation{Department of Mechanical and Aerospace Engineering, Princeton University, Princeton, NJ 08544, USA}

\author{Fran\c{c}ois Boulogne}
\affiliation{Department of Mechanical and Aerospace Engineering, Princeton University, Princeton, NJ 08544, USA}

\author{Eujin Um}
\affiliation{Department of Mechanical and Aerospace Engineering, Princeton University, Princeton, NJ 08544, USA}

\author{Ian Jacobi}
\affiliation{Department of Mechanical and Aerospace Engineering, Princeton University, Princeton, NJ 08544, USA}

\author{Ernie Button}
\affiliation{E. Pedro Road, Phoenix, AZ  85042, USA}
\author{Howard A. Stone}\email{hastone@princeton.edu}
\affiliation{Department of Mechanical and Aerospace Engineering, Princeton University, Princeton, NJ 08544, USA}

\date{\today}

\begin{abstract}
Surface coatings and patterning technologies are essential for various physicochemical applications. In this Letter, we describe key parameters to achieve uniform particle coatings from binary solutions: First, multiple sequential Marangoni flows, set by solute and surfactant simultaneously, prevent non-uniform particle distributions and continuously mix suspended materials during droplet evaporation. Second, we show the importance of particle-surface interactions that can be established by surface-adsorbed macromolecules. To achieve a uniform deposit in a binary mixture, a small concentration of surfactant and surface-adsorbed polymer (0.05 wt\% each) is sufficient, which offers a new physicochemical avenue for control of coatings.
\end{abstract}

\maketitle
An evaporating liquid drop, either single or multi-component, containing solutes or particulates leaves a deposit whose form is determined by various parameters, for instance internal flow fields~\cite{Deegan:1997fk,Hu:2006gp,ristenpart2007influence}, liquid compositions~\cite{Christy:2011jn,Bennacer:2014fz,Park:2006cp,Still:2012bu, sempels2013auto,cazabat2010evaporation, poulard2007control}, and interactions between suspended particles and a solid substrate~\cite{shmuylovich2002surface,Deegan:2000tb,yunker2011suppression,askounis2014effect}, which are crucial for coating processes. In particular, control of the deposit uniformity and thickness can be important in surface patterning~\cite{kuang2014controllable, han2012learning, cai2008marangoni}, ink-jet~\cite{mishra2010high, Park:2006cp, Emma:2015} and 3D printing technologies~\cite{kong20143d}. These processes are complex because of physicochemical dynamics that arise from Marangoni effects~\cite{Deegan:2000tb, Hu:2006gp, Christy:2011jn, Bennacer:2014fz, cazabat2010evaporation, poulard2007control, Still:2012bu, sempels2013auto, majumder2012overcoming, sefiane2014patterns} and particle deposition mechanisms~\cite{Deegan:2000tb,askounis2014effect,shmuylovich2002surface,boulogne2015homogeneous}. In fact, although a binary mixture is used quite often to achieve uniform particle deposition from droplets smaller than 100 $\upmu$m~\cite{mishra2010high, Park:2006cp, Emma:2015}, to our best knowledge such coatings have not been achieved for larger droplets. Furthermore, while the wetting and dewetting behaviors of binary mixture drops have been investigated~\cite{Sefiane:2008ft, sefiane2003experimental}, the relation between the deposition pattern and the evaporatively driven flow field in a binary mixture droplet is incomplete (Table~S1, Supporting Information (SI))~\cite{supportinginformation}.

In this Letter, to achieve a uniform coating, we identify key characteristics of a multicomponent solution, which consists of a binary mixture, surface-active surfactant, and surface-adsorbed polymer. We were motivated to pursue the ideas here from examining a whisky droplet after drying on an ordinary glass where it creates a relatively uniform particle deposit (see Fig. 1), which is in contrast to the well-known `coffee-ring stain'~\cite{Deegan:1997fk}. Based on our understanding of the drying and coating mechanisms of binary liquid droplets, whisky droplets, and more complex solution droplets, we design a model liquid that yields nearly uniform deposits by taking the approach that whisky is an ethanol-water mixture containing diverse dissolved molecules, which contribute to the complexity of the system, the flows, and the final particle deposits.
\begin{figure}[b]\centering
\includegraphics[trim=0.0cm 0.2cm 0.0cm 0.2cm, clip=true, width=0.25\textwidth]{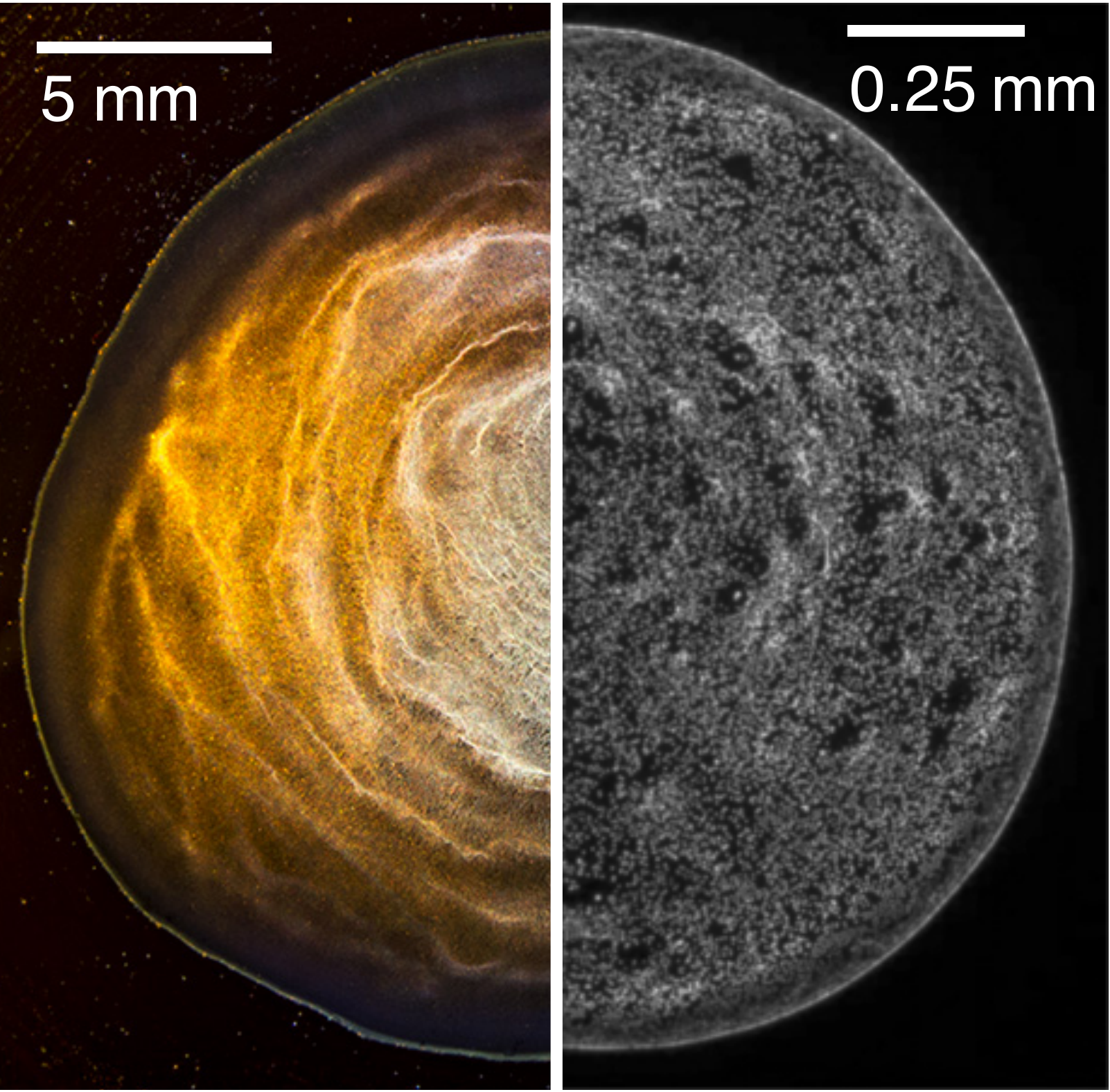}
\caption{Left: A dried mark of a whisky droplet (Macallan, UK) on a normal glass. The image is obtained using an orange color flashlight. Right: A dried deposit pattern of a Glenlivet whisky (UK) with fluorescent polystyrene particles.}
\label{Fig1}
\end{figure}

\begin{figure*}\centering
\includegraphics[trim=0.0cm 0.0cm 0.0cm 0.0cm, clip=true, width=0.90\textwidth]{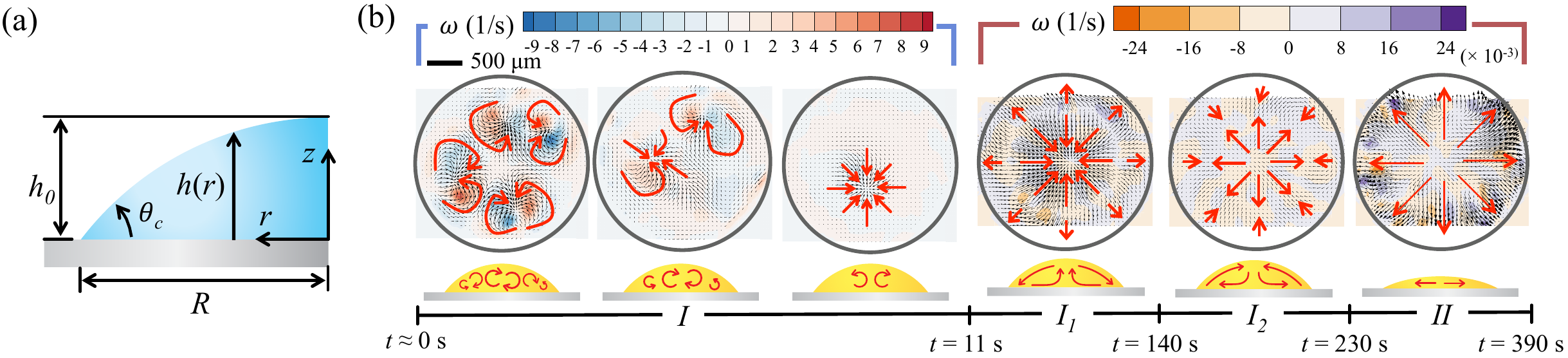}
\caption{(a) Sketch of a liquid drop on a solid substrate. (b) Flow fields (vectors) and wall-normal vorticity $\omega$ fields (color contours) of a Glenlivet whisky. The flow field was measured near the substrate. The total drying time was about 470 s. Below each flow field plot a schematic of the side view of the evaporating droplet is provided. The red arrows represent the flow pattern. There are different flow regimes, multiple vortices (I), two circulatory flows (I$_{1}$ and I$_{2}$), and  radial outward flow (II). At the stage II, from the outward radial flow, we estimate that the ethanol is almost evaporated and there is no significant surfactant effect along the droplet interface.}
\label{Fig2}
\end{figure*}
We begin with a few remarks about whisky, since it serves as a model complex mixture, where nearly uniform particle deposits are observed after drying. Whisky is an alcoholic liquid (ethanol:water, 35:65 \% by weight) made by the hydrolytic breakdown of cereal starches into fermentable sugars and their subsequent fermentation and distillation~\cite{Yoneya2005}; see SI for a brief summary~\cite{supportinginformation}. During the manufacturing procedures, various chemicals are formed, e.g. cellulose, hemicellulose, lignin, and a number of highly extractable molecules, including lipids, acids, sugars, and tannins~\cite{Lee2001}. However, their volume fractions are smaller than 1 \% in total~\cite{Yoneya2005}.

We investigated the flow field inside an evaporating whisky drop by using Particle Image Velocimetry (PIV) and recorded images of the final particle deposits (Movie 1 in~\cite{supportinginformation}). To visualize the flow field inside droplets, we added 1 $\upmu$m diameter fluorescent particles (carboxylate-modified polystyrene, Invitrogen, USA) at a concentration of 8 $\times$ 10$^{-4}$ vol\%. A liquid volume 0.60 $\pm$ 0.07 $\upmu$l was deposited on top of a solid substrate (VWR, USA) (see experimental details, Fig.~S1 and Section S2, SI). During evaporation, the temperature and relative humidity were fixed, i.e. T = 299 K and RH = 50 \%. The whisky drop (Glenlivet, UK) initially has radius $R$ = 1.3 mm, height $h_{0}$ = 0.46 mm, and apparent contact angle $\theta_{c}$ = 36$^{\circ}$ (see notations in Fig. 2(a)).

Initially (regime I), multiple vortices are observed as shown in Fig. 2(b), which is similar to the flow pattern of an ethanol-water (35:65 wt\%) mixture~(Fig.~S2(a), SI). The complicated mixing flows are driven by solutal-Marangoni effects caused by a concentration variation because of the evaporation of ethanol~\cite{Christy:2011jn, Bennacer:2014fz}. Due to this Marangoni flow, the particles are distributed everywhere. The typical flow speed is $U$ = $\mathcal{O}$(100 $\upmu$m/s) and the wall-normal vorticity is $\omega$ = $\left(\frac{\partial u_y}{\partial x} - \frac{\partial u_x}{\partial y}\right)$ = $\mathcal{O}$(1 s$^{-1}$) for in-plane velocity $(u_x, u_y)$.

After regime I, the flow is directed radially outward along the air-liquid interface and radially inward along the substrate (see the schematic side view of regime I$_{1}$ of Fig. 2(b)). The flow speed is $U$ = $\mathcal{O}$(1 $\upmu$m/s) and the vorticity becomes weaker compared to regime I, e.g. $\omega$ = $\mathcal{O}$(10$^{-3}$ s$^{-1}$), as the size and strength of the vortex change. As the whisky drop evaporates further, we observed a reversed flow pattern showing an outward radial flow along the substrate and an inward radial flow along the air-liquid interface (regime I$_{2}$ of Fig. 2(b)). Next, an outward capillary flow is observed as shown in regime II of Fig. 2(b)~\cite{Deegan:1997fk}. Thus, by this time we can assume that ethanol is almost completely evaporated. The distinct particle deposits after whisky completely dries appear linked to the flow fields identified as regimes I$_{1}$ and I$_{2}$, which are not observed in the ethanol-water (35:65 wt\%) mixture droplet (Movie 2 in~\cite{supportinginformation}). Therefore, an ethanol-water mixture can not produce a uniform deposit (Fig.~S2(b), SI).

From the flow field differences between the whisky drop and the ethanol-water mixture drop, we suspect that some chemical compounds play a role in this flow field. To check we completely dried the whisky at room temperature (T = 298 K) and the dried solid residue of whisky was resolubilized in deionized water. Then, we investigated the flow field of this mixture droplet during evaporation. We observed that particles accumulated at the contact line were released from the contact line and moved along the liquid-air interface to the top center of the droplet due to a surfactant-driven Marangoni effect~\cite{marin2015surfactant} (Movie 3 in~\cite{supportinginformation}). We measured the surface tension of this solution to be 60.5 mN/m, which is lower than distilled water's surface tension 72.0 mN/m, and so we conclude that whisky contains molecules acting as surfactants. Natural phospholipids from various grains of whisky's raw materials including barley, wheat, corn, and rye have been detected in whisky and, are the most likely source of these natural surfactants (Section S1, SI).

\begin{figure*}\centering
\includegraphics[trim=0.0cm 0.0cm 0.0cm 0.0cm, clip=true, width=0.65\textwidth]{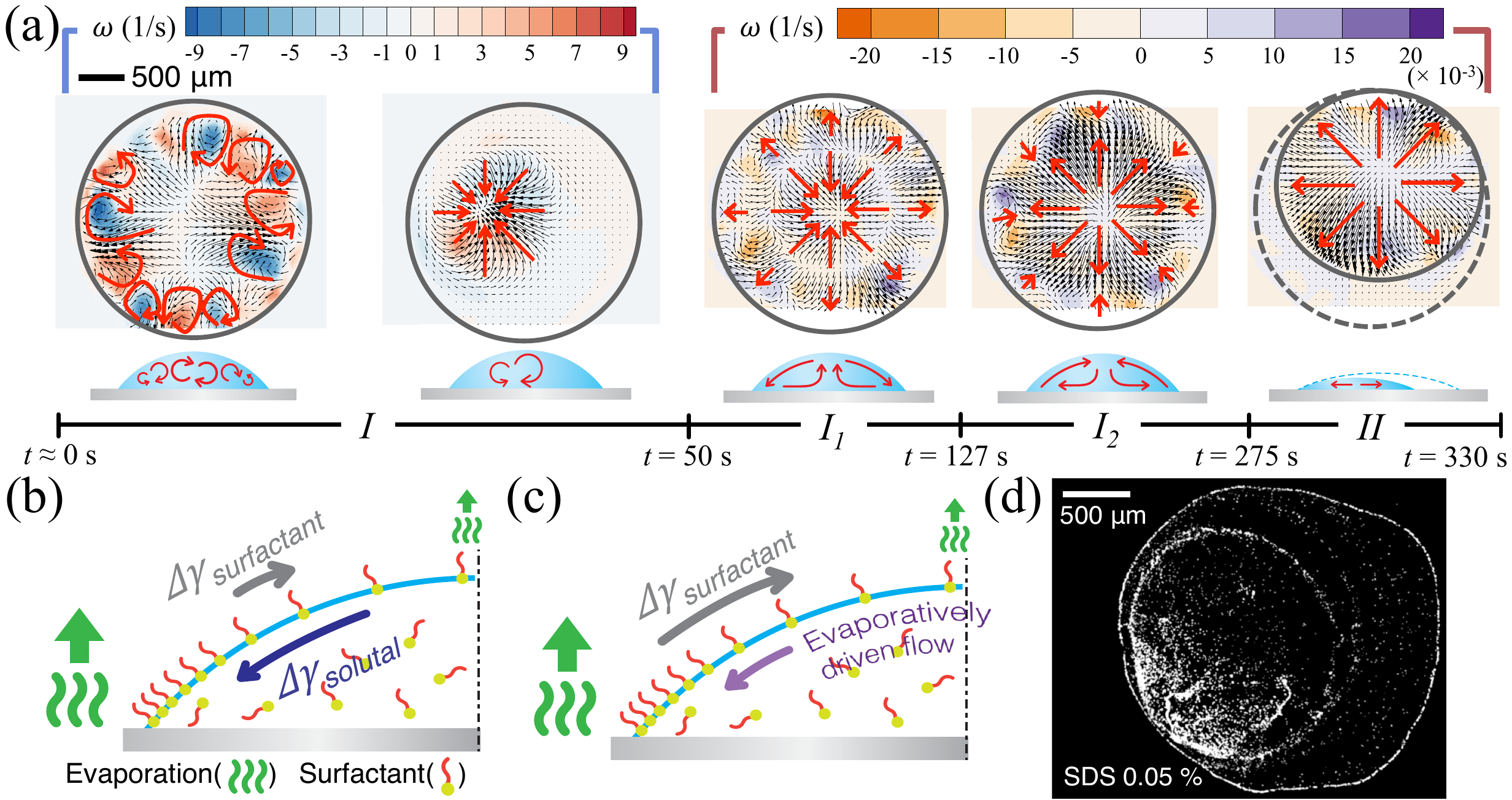}
  \caption{(a) Flow fields (vectors) and wall-normal vorticity $\omega$ fields (color contours) of an ethanol-water (35:65 wt\%) mixture with 0.05 wt\% SDS. Below each flow field plot a schematic of the side view of the evaporating droplet is provided. The red arrows represent the flow pattern. The total drying time was about 400 s. At the stage II, from the outward radial flow, we estimate that the ethanol is almost evaporated and there is no significant surfactant effect along the droplet interface. Schematic of (b) solutal and surfactant-driven Marangoni effects and (c) the surfactant-driven Marangoni effect and the evaporatively driven flow effect along the drop interface. The grey, dark blue, and light purple arrows indicate the surfactant, solutal, and evaporative flux effects, respectively. (d) The final deposition pattern of the binary mixture drop with SDS on the cover glass where the particle concentration is 8 $\times$ 10$^{-4}$ vol\%.}
  \label{Fig3}
\end{figure*}
To check the effect of the surfactant, we prepared an ethanol-water (35:65 wt\%) mixture containing 0.05 wt\% sodium dodecyl sulfate (SDS, Sigma-Aldrich, USA), and we recorded the flow field and dried pattern. SDS is a common surfactant that has been used in previous studies of Marangoni flows~\cite{Deegan:2000tb,Still:2012bu,marin2015surfactant}. In our system, the initial surfactant concentration is lower than the critical micelle concentration~\cite{suzuki1976studies}. By adding surfactants, we mimicked the flow pattern of a drying whisky drop (Fig. 2(b) and 3(a)): two different circulating flows are observed after the initial multiple vortical flows. As the droplet evaporates, the ethanol concentration near the contact line is lower than that of the drop center due to the non-uniform evaporative flux along the droplet height~\cite{Bennacer:2014fz}, so the solutal Marangoni stress occurs along the droplet interface (Fig. 3(b)). Simultaneously, as the surface-active molecules, which in this case are a dissolved surfactant, accumulate at the contact line~\cite{Still:2012bu} and the ethanol concentration decreases in time, the surfactant-driven Marangoni stress becomes dominant (Fig. 3(c)). This flow transition indicates that initially a solutal Marangoni effect is dominant compared to a surfactant-driven Marangoni effect. From this and based on the literature~\cite{Bennacer:2014fz, cazabat2010evaporation,marin2015surfactant}, in this problem we can establish the hierarchy of Marangoni effects, i.e. 1) solutal, 2) surfactant, and 3) the thermal Marangoni effect. As a consequence of this competition between solutal- and surfactant-driven Marangoni stresses, we observe the sequence of opposite signed circulatory flows.
The critical temporal evolution of the circulation transition can be investigated further by studying the droplet shape or the concentration of solute and surfactant.
At longer times, if the surfactant is saturated everywhere, a typical outward radial capillary flow is observed (regime II, Fig.~2(b) and 3(a)).

The critical condition to induce the Marangoni flow caused by a surfactant and/or solute is considered next. The ethanol-water droplet with SDS is thin because $h_{0}/R$ $<$ 1 where $h_{0}$ $\approx$ 100 $\upmu$m and $R$ $\approx$ 1 mm for late times (e.g. after regime I). The typical flow speed $U$ $\approx$ 1 $\upmu$m/s (from PIV results), so that the Reynolds number $\mathrm{Re} = \rho Uh_{0}/\mu$ $\approx$ 10$^{-4}$, where density $\rho$ $\approx$ 10$^{3}$ kg/m$^{3}$ and viscosity $\mu$ $\approx$ 1 mPa$\cdot$s. Furthermore, the surface tension force is dominant compared to both the viscous and gravity forces, as the capillary number $\mathrm{Ca}=\mu U/\gamma$ $\approx$ 10$^{-7}$ and the Bond number $\mathrm{Bo}=\rho g R^2/\gamma$ $\approx$ 10$^{-1}$, where $g$ = 9.8 m/s$^{2}$ is gravity and $\gamma$ $\approx$ 72 mN/m is the surface tension of water. Therefore, by using the lubrication approximation, in cylindrical ($r, z$) coordinates the Navier-Stokes equations can be simplified and the surface velocity $u(r, t)$ at the liquid-air interface $z = h(r, t)$, nearly, a spherical cap, can be expressed as (see details in Section S4, SI)
\begin{equation}\label{surface_velocity}
u(r,t) = \underbrace{\frac{\gamma h^{2}}{2\mu}\left(\frac{\partial p}{\partial r} \right)}_{\text{capillary effect}}+\overbrace{\frac{h}{\mu}\frac{\partial \gamma}{\partial r}}^{\text{Marangoni effect}} ~\mbox{at} ~z = h(r, t).
\end{equation}
Here, the capillary pressure $p$ = -$\gamma\nabla^{2}\tilde{h}$ where $\tilde{h}$($\ll h_{0}$) is the perturbation to the liquid-air interface caused by the internal flow. Then, the interfacial velocity driven by the capillary pressure gradient scales as $(\gamma h_{0}^{2}\tilde{h})/(\mu R^{3})$ and the interfacial velocity driven by Marangoni effects is expected to have a magnitude of $(h_{0}\Delta\gamma)/(\mu R)$ where the sign of $\Delta\gamma$ determines the flow direction. If both velocities have the same order of magnitude,
\begin{equation}\label{velocity}
 \left| \frac{\Delta\gamma}{\gamma}\frac{R^2}{h_{0}\tilde{h}} \right| ~\approx~ 1,
\end{equation}
then, for large interface deformation $\tilde{h}$ $\rightarrow h_{0}$, we obtain the upper bound for $|\Delta\gamma|$ of about 1 mN/m. This value is consistent with previous studies on the interface deformation of an evaporating droplet by Marangoni effects~\cite{kajiya2009controlling, poulard2007control}. On the other hand, experimental observations indicate $|\Delta\gamma|$ $\approx$ 1 $\upmu$N/m, for an evaporating water drop with SDS, which maintains a nearly spherical cap shape~\cite{marin2015surfactant}. From Eq.~(\ref{velocity}), we estimate $\tilde{h}$ $\sim$ 0.1 $\upmu$m, which is negligible compared to the droplet size.

Although we mimicked the flow pattern of the drying of a whisky drop by adding SDS to a binary mixture, the particles are not uniformly distributed on the substrate, as shown in Fig. 3(d). We observed that in the model liquid drop when the contact line recedes, the contact line transports particles towards the center of the drop (Movie 4 in~\cite{supportinginformation})~\cite{Deegan:2000tb, berteloot2008evaporation}. However, for a whisky drop, although the contact line recedes, the particles remain nearly uniformly distributed on a substrate (Movie 1 in~\cite{supportinginformation}).

The chemical composition of whisky has been extensively investigated. According to the literature (see Section S1, SI), whisky contains natural polymers (e.g. lignin and polysaccharides). We hypothesize that some macromolecules, originally present in whisky, adsorb on a substrate and may play a role in adhesion and retention of the particles on the substrate. To test this idea, we added polymer, polyethylene oxide (PEO) (0.05 wt\%), to the ethanol-water mixture with surfactant (0.05 wt\%). At this polymer concentration, the polymer does not influence the flow field until regime I$_{2}$. When the contact line recedes (regime II), the added polymer contributes to capture the particles on the surface but without polymers the receding contact line transports particles (see Movies 4 and 5 in~\cite{supportinginformation}).

It is known that PEO can adsorb onto silica~\cite{braithwaite1997effect, trens1993conformation, de1987polymers} creating a ``pseudo-brush'' structure on the glass surface. The spatial density of adsorbed polymer is about 1 mg/m$^{2}$~\cite{cabane1997shear}, such that the quantity of adsorbed polymer is extremely small compared to the suspended polymer. As evaporation proceeds, the polymer concentration in the droplet increases. The polymer adheres on the silica substrate, which is not transported by the receding contact line. As a result, the particles are captured by a dense polymer structure and then remain adhered on the substrate. This adherence mechanism can be reproduced with different molecular weights (2 $\times$ 10$^{4}$ -- 4 $\times$ 10$^{6}$ Da, PEO) and other polymers, e.g. hydroxyethycellulose, polyvinyl alcohol, and polyvinylpyrrolidone (Fig.~S3, SI).

Also, we tested the effect of polymer without surfactant and a primary ring pattern is observed along the contact line (see Fig.~4), which is the signature of the coffee ring effect. As a result, the surfactant is crucial to prevent particle accumulation along the contact line.
\begin{figure}
  \centering
  \includegraphics[trim=0.0cm 0cm 0.0cm 0cm, clip=true, width=0.347\textwidth]{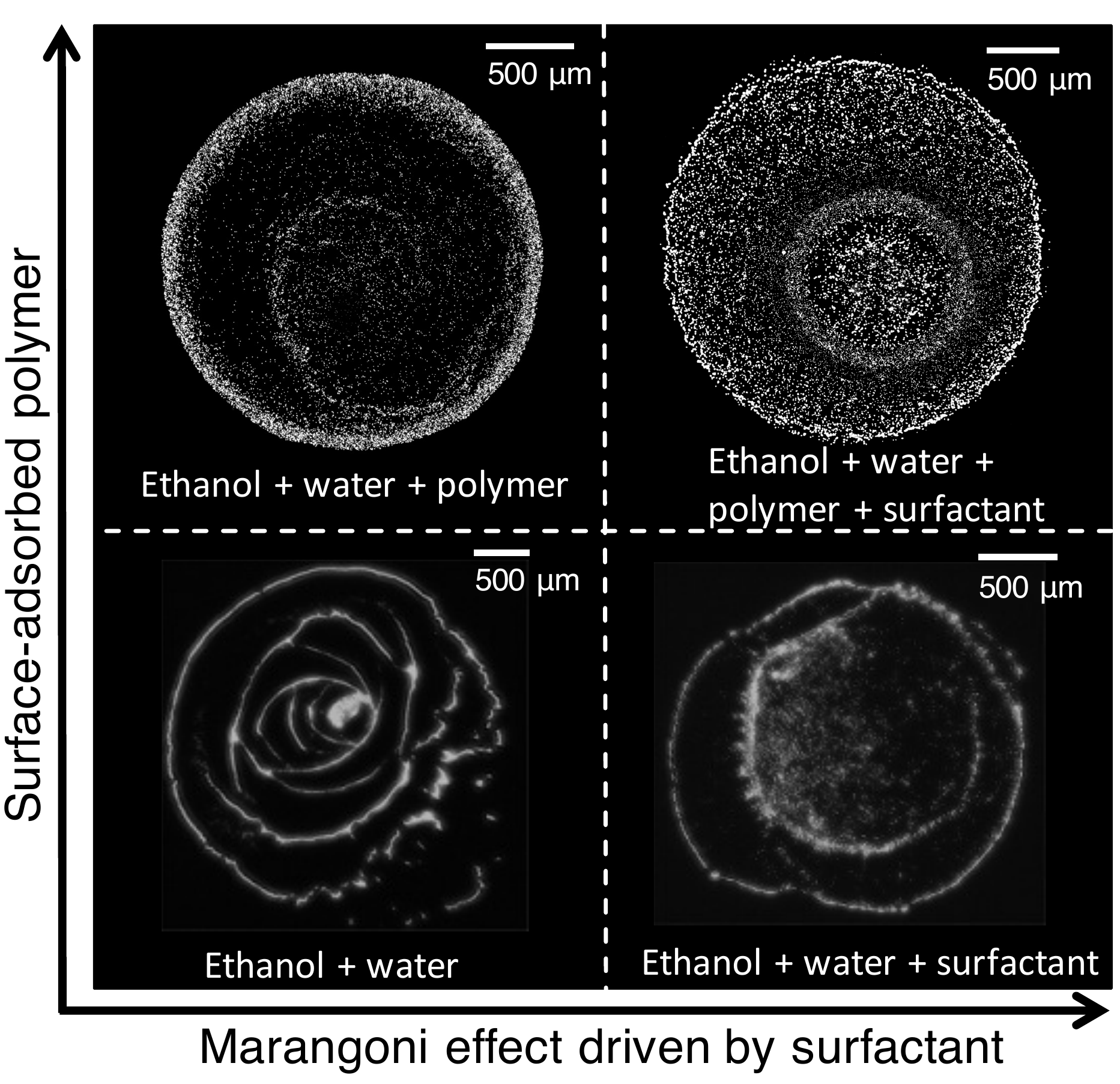}
    \caption{A diagram of effects of surfactant driven Marangoni flows and surface-adsorbed materials in the binary mixture on the final deposit. The concentration of ethanol is 35 wt\% in DI water. PEO (4 $\times$ 10$^{6}$ Da) and SDS concentration are 0.05 wt\%, respectively.}
  \label{Fig4}
\end{figure}

In Fig. 5(a), we compare the final deposition pattern of whisky, water, and the model liquid (respectively, from left to right), which are deposited on top of a cover glass. We then measure the average particle number density as a function of radial location (Fig. 5(b)), which exhibits significant correspondence in coating uniformity between whisky and the model liquid. As shown in Fig. 5, the proposed model liquid can produce a nearly uniform deposit. Here, we should note that different types of polymer can create different patterns (Fig.~S3(b-f), SI). Presumably, the surface adsorbed macromolecules in whisky are not identical with the polymers that we used in this study. We also obtained a nearly uniform particle deposition pattern with another glass substrate, which has a lower contact angle with water (Fig.~S4, SI).
\begin{figure}
  \centering
  \includegraphics[trim=0.0cm 0cm 0.0cm 0cm, clip=true, width=0.41\textwidth]{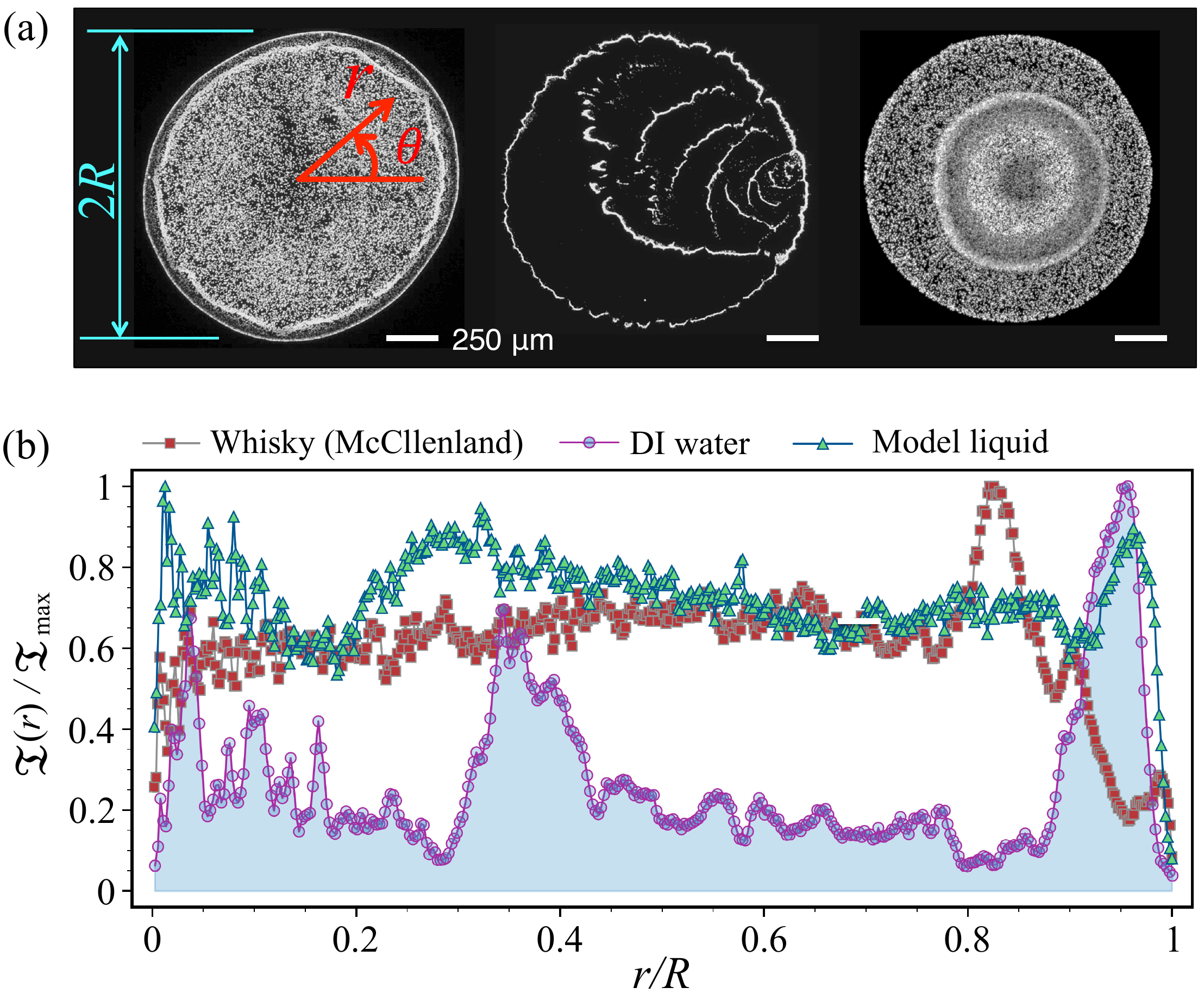}
    \caption{(a) Comparison of the final deposition patterns on top of the cover glass (VWR, USA): (Left) whisky (McCllenland, UK), (Center) water, and (Right) a model liquid (ethanol:water (35\% : 65\%) + SDS (0.05\%) + PEO (0.05\%) by weight) containing 1 $\upmu$m polystyrene particles (5 $\times$ 10$^{-3}$ vol\%). 2$R$ is the diameter of the final area. (b) Deposit profiles along $r$ are plotted for each of the images in (a). The intensity profile $\mathfrak{T}(r)$ is normalized with the maximum of $\mathfrak{T}(r)$ where $\mathfrak{T}(r)$ = $\frac{1}{2\pi}$$\int_{0}^{2\pi} i(r,\theta) d\theta$ and $i(r,\theta)$ is the local light intensity.}
  \label{Fig5}
\end{figure}

In this Letter, we have shown that a combination of a binary mixture, surfactant, and surface-adsorbed polymer influences the final deposition pattern so that more uniform deposits occur.
Based on understanding of the drying and coating mechanisms, we demonstrate that continuous mixing by distinct Marangoni flows and strong interaction between particles and a substrate are important to obtain a uniform deposit. Although the complex chemistry of whisky is not fully understood, we believe that these observations inspired by whisky drying are useful and applicable to coating processes more generally and the proposed method offers a new physicochemical avenue for control of coatings. To accomplish more controlled coatings, a future study can be devoted to analyze the interplay between the flow characteristics and the microstructure of a final deposit with respect to the spatial~\cite{Deegan:1997fk} and temporal~\cite{marin2011order} variations.

H.K. and H.A.S. acknowledge the DOE for support via grant no. DE-SC0008598. F.B. acknowledges that the research leading to these results received funding from the People Programme (Marie Curie Actions) of the European Union's Seventh Framework Programme (FP7/2007-2013) under REA grant agreement 623541. We thank A. Perazzo, C. Poulard, E. Rio, H. Gelderblom, J. Nunes, and Y.L. Kong for helpful discussions.

\newpage 

\begin{center}
\Large Supporting Information
\end{center}

\beginsupplement
     \renewcommand*{\citenumfont}[1]{S#1}
\renewcommand*{\bibnumfmt}[1]{[S#1]}


\section*{S1. Whisky manufacturing procedures}

The raw cereals (e.g. barley, wheat, corn, or rye) are first broken down into sugars through enzymatic action in a process known as mashing. This process also draws nitrogen-rich amino acids into solution, which will eventually provide food for yeast growth. The result of the mashing (called `wort')  is inoculated with yeast to begin the fermentation process, which, over the course of a 2--3 days, produces a variety of important flavor molecules, including organic acids, higher alcohols, esters, ketones, and aldehydes, all of which are sufficiently volatile to be preserved throughout the final distillation \cite{Hui2012}.

The resulting `wash' from fermentation is then distilled (in batches for malt whiskies or continuously for grain whiskies). The heat of distillation drives reactions which produce another family of flavor molecules, in addition to those generated during fermentation, and which are all collectively known as congeners. Unlike highly purified vodka, these congeners are retained in the final whisky product and are known for producing a distinctive flavor profile unique to each whisky recipe \cite{Aylott1994} and measurable using gas chromatography \cite{Duncan1966,Kahn1968,Williams1972,Salo1972}.

Fresh from distillation, the whisky is practically colorless with very harsh flavors, and thus the final step in whisky production, which distinguishes it from other distilled liquor, is  maturation of the whisky in wooden casks. American or European white oak barrels are first charred (at 473 K) in order to produce a chemically active surface \cite{Clyne1993} and then filled with diluted distillate. The oak is made up of cellulose, hemicellulose, lignin, and a number of highly extractable molecules, including lipids, acids, sugars, and tannins \cite{Lee2001}. During the aging process, the distillate undergoes internal reactions with oxygen, which can permeate through the casks, as well as reactions with the surrounding wood, and finally the distillate extracts molecules from the wood itself \cite{Piggot1993}.

The extracted molecules provide some of the amber color associated with whisky (although most is likely due to artificial addition of caramel \cite{Yoneya2005}) and have a significant effect on the final flavor and aroma. Lignin-related compounds are the most significant extract from the wood casks \cite{Conner1992} and serve to displace other volatile components from the air-whisky interface upon consumption, thereby drastically altering the so-called `headspace' of odor directly above the whisky glass. Other wood extracts can also significantly affect this flavor release \cite{Conner1999}. Many of these wood extracts, like lipids, fatty acids, lactones, and phenols, can also affect the clarity of the whisky.

\section*{S2. Experimental details}
For flow visualization, we performed Particle Image Velocimetry (PIV). We added 1~$\upmu$m carboxylate-modified polystyrene fluorescent particles into the working fluid. The Rhodamine-B fluorescent particles were illuminated with a green light source ($\lambda$ = 520 nm). The fluorescent signals from the particles ($\lambda$ $>$ 540 nm) were captured by either a high-speed CMOS camera (Phantom v7.3, USA) having a pixel resolution of 800 $\times$ 600 and a 8-bit dynamic range at a frame rate of 100 fps or a CCD camera built in a Leica microscope (Leica DFC 300 FX, Wetzlar, Germany) having a resolution of 1392 $\times$ 1040 pixels and a 12-bit dynamic range at a frame rate of 5 fps.
\begin{table*}
\caption{Studies on particle deposition patterns of an evaporating droplet. ($\bigcirc$: yes, $\times$: no, and $\bigtriangleup$: no quantitative measurement)} 
\label{refs-tab}\centering
\begin{tabular}{p{5cm} c c c c c c}
\hline
References & Flow field & Deposition & Binary liquid  & Surfactant \\
&  measurement & pattern & mixture & \\ 
\hline
Deegan et al. (1997)~\cite{Deegan:1997fk-SI},\newline Shmuylovich et al. (2002)~\cite{shmuylovich2002surface-SI} & $\bigtriangleup$ & $\bigcirc$ & $\times$ & $\times$ \\
& & & & \\
Deegan (2000)~\cite{Deegan:2000tb-SI} & $\times$ & $\bigcirc$ & $\times$ & $\bigcirc$\\
& & & & \\
Christy et al. (2011)~\cite{Christy:2011jn-SI}, \newline Bennacer and Sefiane (2014)~\cite{Bennacer:2014fz-SI} & $\bigcirc$ & $\times$ & $\bigcirc$ & $\times$\\
& & & & \\
Park and Moon (2006)~\cite{Park:2006cp-SI}  & $\times$ & $\bigcirc$ & $\bigcirc$ & $\times$ \\
& & & & \\
Majumder et al. (2012)~\cite{majumder2012overcoming-SI} & $\times$ & $\bigcirc$ & $\bigcirc$ & $\times$\\
& & & & \\
Askounis et al. (2014)~\cite{askounis2014effect-SI} & $\bigtriangleup$ & $\bigcirc$ & $\bigcirc$ & $\times$\\
& & & & \\
Still et al. (2012)~\cite{Still:2012bu-SI} & $\bigtriangleup$ & $\bigcirc$ & $\times$ & $\bigcirc$\\
& & & & \\
Sempels et al. (2013)~\cite{sempels2013auto-SI}  & $\bigtriangleup$ & $\bigcirc$ & $\times$ & $\bigcirc$ \\
& & & & \\
Marin et al. (2016)~\cite{marin2015surfactant}  & $\bigcirc$ & $\times$ & $\times$ & $\bigcirc$ \\
& & & & \\
Current work   & $\bigcirc$ & $\bigcirc$ & $\bigcirc$ & $\bigcirc$ \\
\hline
\end{tabular}
\end{table*}

The flow field near the substrate was measured using PIV. We obtained in-plane velocity vectors, and the wall-normal vorticity was calculated based on the velocity vector field. The velocity vector calculation was performed using an open-source software, PIVlab, which is a time-resolved digital particle image velocimetry tool for Matlab~\cite{SIthielicke2010pivlab}. For the high-speed measurements, the velocity was obtained by performing iterative 2D cross-correlations of the particle distribution with multiple interrogation windows of 64 $\times$ 64 pixels with 50\% overlap for the coarse grid and 32 $\times$ 32 pixels with 50\% overlap for the refined grid system. Also, for recording with the microscope CCD camera, iterative 2D cross-correlations were performed with multiple interrogation window sizes. The first interrogation window used was 128 $\times$ 128 pixels with 50\% overlap and then the next interrogation window used was 64 $\times$ 64 pixels with 50\% overlap. The random error of the PIV measurements is about 0.02 pixel units for given interrogation domains. This error is consistent with typical measurement uncertainty~\cite{SIadrian2011particle}.

During the experiments, to control the temperature and relative humidity, we designed thermo-hygro-control hood. The evaporating droplet was kept in the control hood. The relative humidity is controlled by a humidity controller (ETS Microcontroller Model-5100, Glenside, PA, USA). During the experiment, the temperature is set at T = 299.0 $\pm$ 0.6 K and the relative humidity is set at 50 $\pm$ 2 \%.

For the substrate preparation, cover (VWR, USA) and slide (Thermo Scientific, USA) glasses were cleaned in an acetone solution in an ultrasonic bath for 5 min, rinsed with deionized water (Millipore MilliQ, USA) and dried using nitrogen gas. The substrate was kept in an oven for 30 min at 333 K. Apparent contact angles with deionized water on top of the cover and slide glasses were 52$^{\circ}$ and 41$^{\circ}$, respectively.

For the side view measurements, a Nikon D5100 camera with Af-s Nikkor 70-200 mm f/4G lens was installed to observe the side view of an evaporating droplet.
We recorded the side view of a drop deposited on a cover glass. The initial apparent contact angles of the ethanol-water mixture (35\% : 65\% by weight, respectively) was about $34^{\circ}$ and the whisky drop (Glenlivet 12 years old, UK) has $\theta_{c}$ = $36^{\circ}$. The contact angle was determined from a fit to the liquid interface data within a distance of 50 $\upmu$m from the corner tip of the droplet.

\section*{S3. Measurement of physical properties} 
The physical property measurements were performed at T = 298 K. (1) Surface tension measurements were performed by means of a pendant droplet method. 
To calculate the surface tension, we used an in-house Matlab code that is based on the algorithm of Rotenberg et al. (1983)~\cite{SIrotenberg1983determination}. 
We validated the code by comparing with experimental results with a conventional Goniometer (Theta Lite, Biolin Scientific).
(2) The viscosity of all liquids was measured with a Rheometer (Anton-Paar MCR 301, USA) with a sandblasted cylinder system (CC27 geometry).
(3) The weight was measured by a Mettler Toledo XS105 scale.

We purchased several whisky products from a local liquor store, e.g. Glenlivet 12 and 15 years old (UK), Macallan (UK), McClelland(UK), and Glenffidich (UK). In this paper, we present the results of Glenlivet 12 years old and McClelland. The other products have similar results. The McClelland's single malt scotch whisky has $\rho$ = 0.961 g/cm$^{3}$, $\mu$ = 2.24 mPa$\cdot$s, and $\gamma$ = 30.7 mN/m and Glenlivet 12 years old has $\rho$ = 0.960 g/cm$^{3}$, $\mu$ = 2.25 mPa$\cdot$s, and $\gamma$ = 38.6 mN/m. Ethanol (200 proof, anhydrous, 99.5\%, Sigma-Aldrich, USA) was mixed with deionized water. The ethanol-water (35:65 wt\%) mixture has $\rho$ = 0.929 g/cm$^{3}$, $\mu$ = 2.24 mPa$\cdot$s, and $\gamma$ = 31.5 mN/m. The sodium dodecyl sulfate (SDS) (0.05 wt\%) added mixture as $\rho$ = 0.929 g/cm$^{3}$, $\mu$ = 2.24 mPa$\cdot$s, and $\gamma$ = 31.2 mN/m. The final model liquid consists of the binary mixture of ethanol and water, SDS (0.05 wt\%), and polyethylene oxide (PEO) (0.05 wt\%) ($\rho$ = 0.929 g/cm$^{3}$, $\mu$ = 2.23 mPa$\cdot$s, and $\gamma$ = 31.2 mN/m). The SDS (Sigma-Aldrich, USA) was tested and added into the final mixture with a concentration (0.05 -- 0.3 wt\%). Furthermore, we used different polymers, e.g. PEOs (6 $\times$ 10$^{5}$ and 4 $\times$ 10$^{6}$ Da), polyvinyl alcohol (3.1 $\times$ 10$^{4}$ -- 5 $\times$ 10$^{4}$ Da), and polyethylene glycol (2 $\times$ 10${^4}$ and 10$^{5}$) from Sigma-Aldrich in USA, and polyvinylpyrrolidone (2 $\times$ 10$^{4}$ Da) from Calbiochem in USA.
\begin{table}
\caption{Surface tension of the ethanol-water (35\% : 65\% by weight) binary liquid mixture measured for the different surfactant concentrations at $T = 298$ K and 50 \% relative humidity.} 
\label{refs-tab2}\centering
\begin{tabular}{c c c c }
\hline
& SDS concentration & Surface tension & \\
& (wt\%) & (mN/m) & \\
\hline
& 0.05 & 31.2 & \\
& 0.1  & 31.1 & \\
& 0.15 & 31.0 & \\
& 0.3 & 30.8 & \\
\hline
\end{tabular}
\end{table}

\section*{S4. Estimation of Marangoni effect}
We consider an evaporating droplet with radius $R$ and height $h_{0}$. The liquid-air interface is described by the function of $h(r, t)$, which is a nearly spherical cap. Assuming the lubrication approximation $(h_{0}\ll R)$, the Navier-Stokes equations reduces in cylindrical coordinates to
\begin{equation}\label{S1eqn}
\frac{dp}{dr} = \mu \frac{\partial^2 u}{\partial z^2},
\end{equation}
where $p$ is the pressure, $\mu$ the dynamic viscosity, and $u$ the radial velocity. By integrating (\ref{S1eqn}) according to $z$, we can obtain the velocity profile as
\begin{equation}\label{S2eqn}
u(r,z,t) = \frac{1}{\mu}\frac{\partial p}{\partial r}\left(\frac{1}{2}z^{2}-h(r,t)z\right) + \frac{z}{\mu}\frac{\partial \gamma}{\partial r},
\end{equation}
where there is a Marangoni shear stress boundary condition $\left(\frac{\partial \gamma}{\partial r} \neq 0 \right)$ at $z$ = $h$($r$, $t$) and a no slip boundary condition of the liquid-solid interface. Here, the pressure $p = -\gamma\nabla^{2}\tilde{h}$ where $\tilde{h}$ is the perturbation to the liquid-air interface caused by the internal flow. Based on this, we could obtain the velocity at the liquid-air interface $z = h$, such as
\begin{equation}\label{S3eqn}
u(r,z,t) = -\frac{h^{2}}{2\mu}\frac{\partial}{\partial r}\left(\gamma\nabla^{2}\tilde{h}\right) + \frac{h}{\mu}\frac{\partial \gamma}{\partial r}.
\end{equation}
Then, the interfacial velocity driven by the capillary pressure gradient can be estimated as magnitude $(\gamma h_{0}^{2}\tilde{h})/(\mu R^{3})$ and the interfacial velocity driven by Marangoni effects is expected to have magnitude $(h_{0}\Delta\gamma)/(\mu R)$ where the sign of $\Delta\gamma$ determines the flow direction. If both velocities have the same order of magnitude,
\begin{equation}\label{velocity}
 \left| \frac{\Delta\gamma}{\gamma}\frac{R^2}{h_{0}\tilde{h}} \right| ~\approx~ 1.
\end{equation}

\section*{S5. Image analysis and post processing}
All the image analyses and post processing were done with Matlab 2014a. 
For the image processing, we removed the out-of-focus blurred particle images by using a high-pass filter and subtracting background noise, and then a 3 $\times$ 3 Gaussian smoothing was applied to reduce noise of the particle images.
For the post processing, we calculated a vorticity field based on the velocity measurement results where \textbf{$\omega$} = $\nabla$ $\times$ $\textbf{u}$ and $\nabla$ $\equiv$ $(\partial/\partial x, \partial/\partial y)$.
Therefore, the wall-normal vorticity is $\omega$ = $\frac{\partial u_y}{\partial x}$ - $\frac{\partial u_x}{\partial y}$ for the in-plane velocity $(u_x, u_y)$.

For the final deposition pattern of the particles, the deposit profile was averaged along the angular direction, i.e. $\mathbb{I}(r)$ = $\frac{1}{2\pi}$$\int_{0}^{2\pi} i(r,\theta) d\theta$ and $i(r,\theta)$ was a local intensity. 
The intensity profile $\mathbb{I}(r)$ was normalized with the maximum of $\mathbb{I}(r)$.
The radial distribution was normalized with the drop radius $R$, which is the smallest radius enclosing the deposit area.

{}

\begin{figure*}[t]
  \centering
  \includegraphics[trim=0.0cm 0.0cm 0.0cm 0.0cm, clip=true, width=0.85\textwidth]{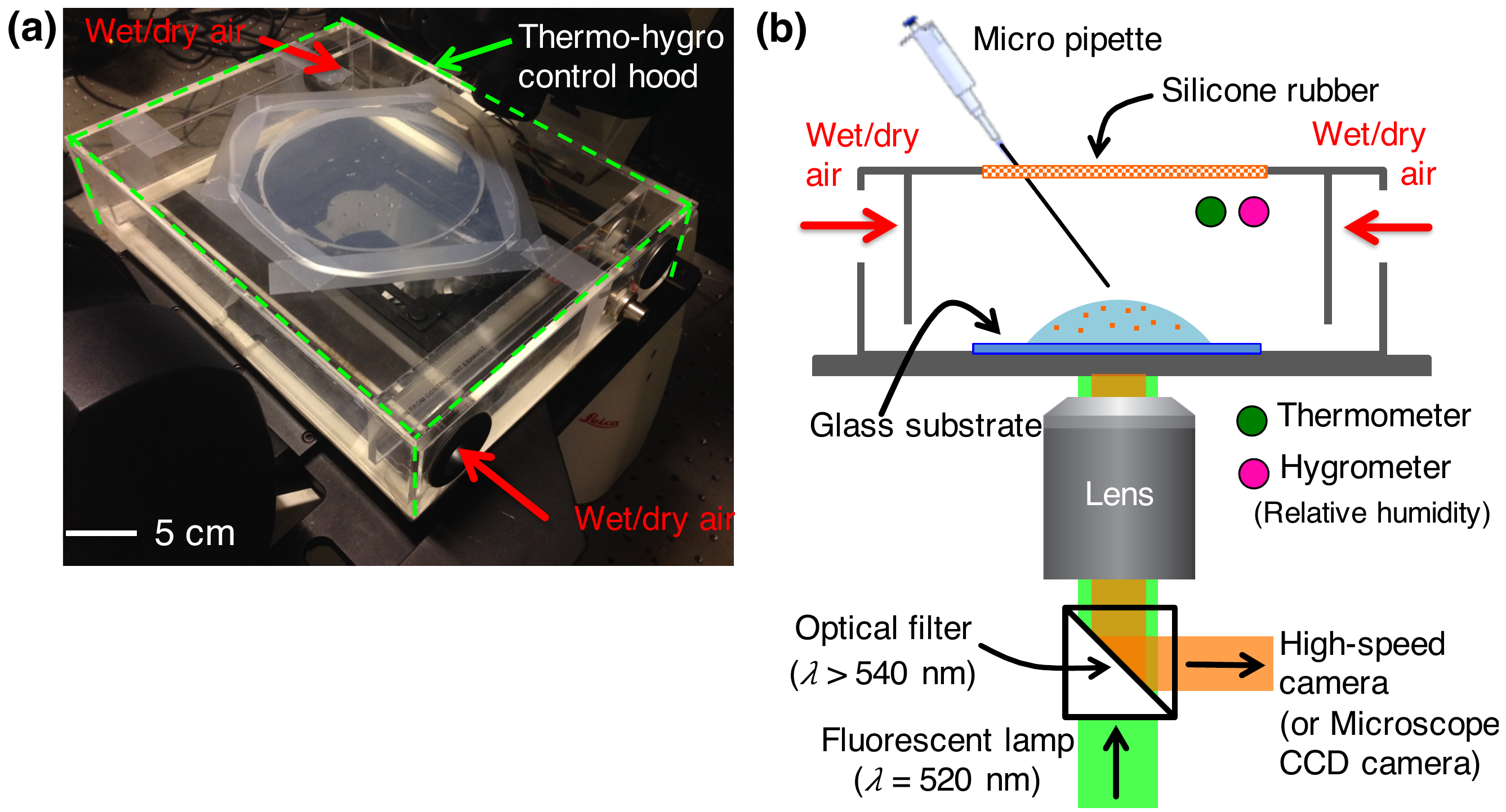}
  \caption{Experimental apparatus. (a) Photograph of the experimental setup on top of an inverted microscope. (b) Schematic of the experimental setup. The relative humidity is controlled by supplying wet and dry air. To measure the movements of Rhodamine-B labeled fluorescent particles, the experiment is illuminated with a green light ($\lambda$ = 520 nm) and a fluorescent particle signal ($\lambda$ $>$ 540 nm) is captured by a high-speed camera or a microscope CCD camera.}
  \label{pivsetup}
\end{figure*}

\begin{figure*}
  \centering
  \includegraphics[trim = 0cm 0cm 0cm 0cm, clip=true, width=0.85\textwidth]{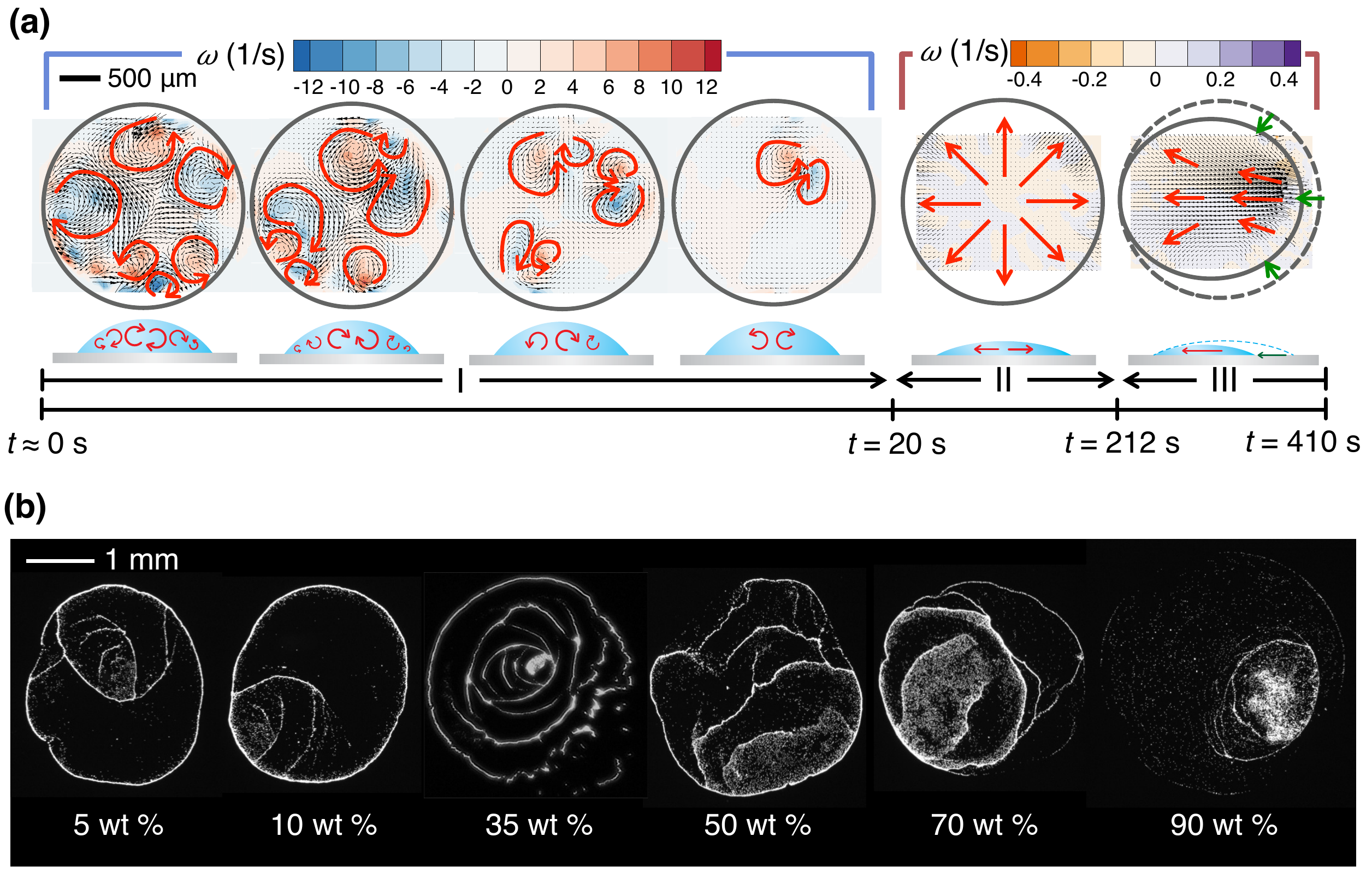}
  \caption{(a) Flow field measurements of an ethanol-water (35:65 wt\%) mixture; the flow (vectors) and wall-normal vorticity $\omega$ (color contours) field. Below each flow field plot a schematic of the side view of the evaporating droplet is provided. The red and green arrows represent the flow pattern and contact line motion, respectively. The flow field was measured near the substrate. The total drying time was about 410 s. There are three regimes, multiple vortices (I), radially outward flow (II) and dewetting flow (III). (b) Final particles deposition patterns for different initial concentrations of ethanol by weight where the particle concentration is 8 $\times$ 10$^{-4}$ vol\%.}
  \label{ethanol_free_whisky}
\end{figure*}

\begin{figure*}
  \centering
  \includegraphics[trim= 0cm 0.0cm 0cm 0.0cm, clip=true, width=0.65\textwidth]{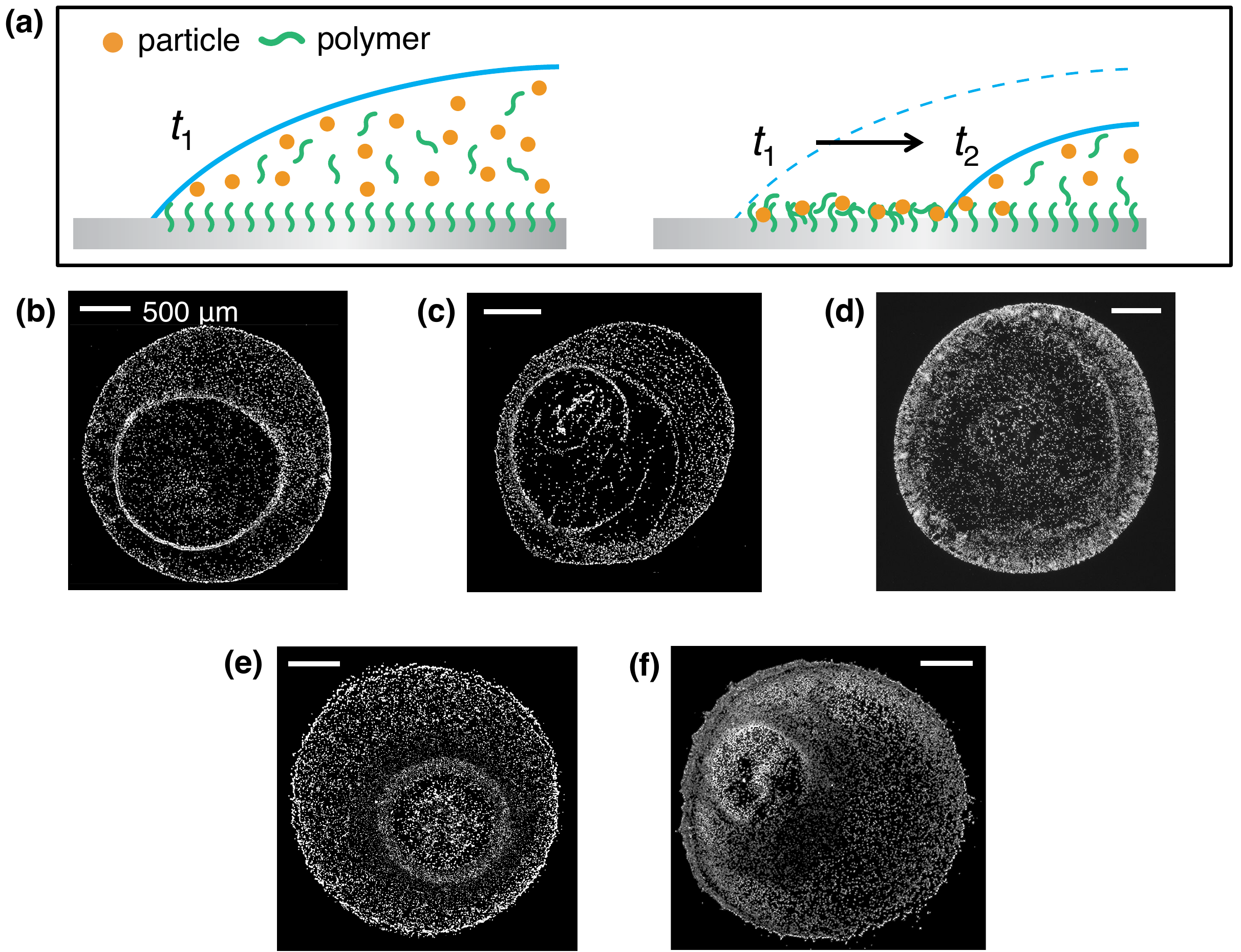}
  \caption{(a) Schematic of the effect of a polymer brush structure on a solid substrate where times $t_1$ $<$ $t_2$. The surfactants are not presented in this schematic. The final deposition pattern of the ethanol-water (35:65 wt\%) mixture + SDS (0.05 wt\%) + a different polymer on the cover glass (VWR, USA). We used (b) polyethylene oxide (4 $\times$ 10$^{6}$ Da), (c) polyethylene oxide (6 $\times$ 10$^{5}$ Da), (d) polyethylene glycol (10$^{5}$ Da), (e) polyvinyl alcohol (3.1 -- 5 $\times$ 10$^{4}$ Da), and (f) polyethylene glycol (2 $\times$ 10$^{4}$ Da). Each polymer (0.05 wt \%) is added into the ethanol-water mixture drop with SDS surfactant. The particle concentration is 8 $\times$ 10$^{-4}$ vol\%. The scale bars represent 500 $\upmu$m.}
  \label{mixture-polymer-flow}
\end{figure*}

\begin{figure*}
  \centering
  \includegraphics[trim=0cm 0cm 0cm 0cm, clip=true, width=0.65\textwidth]{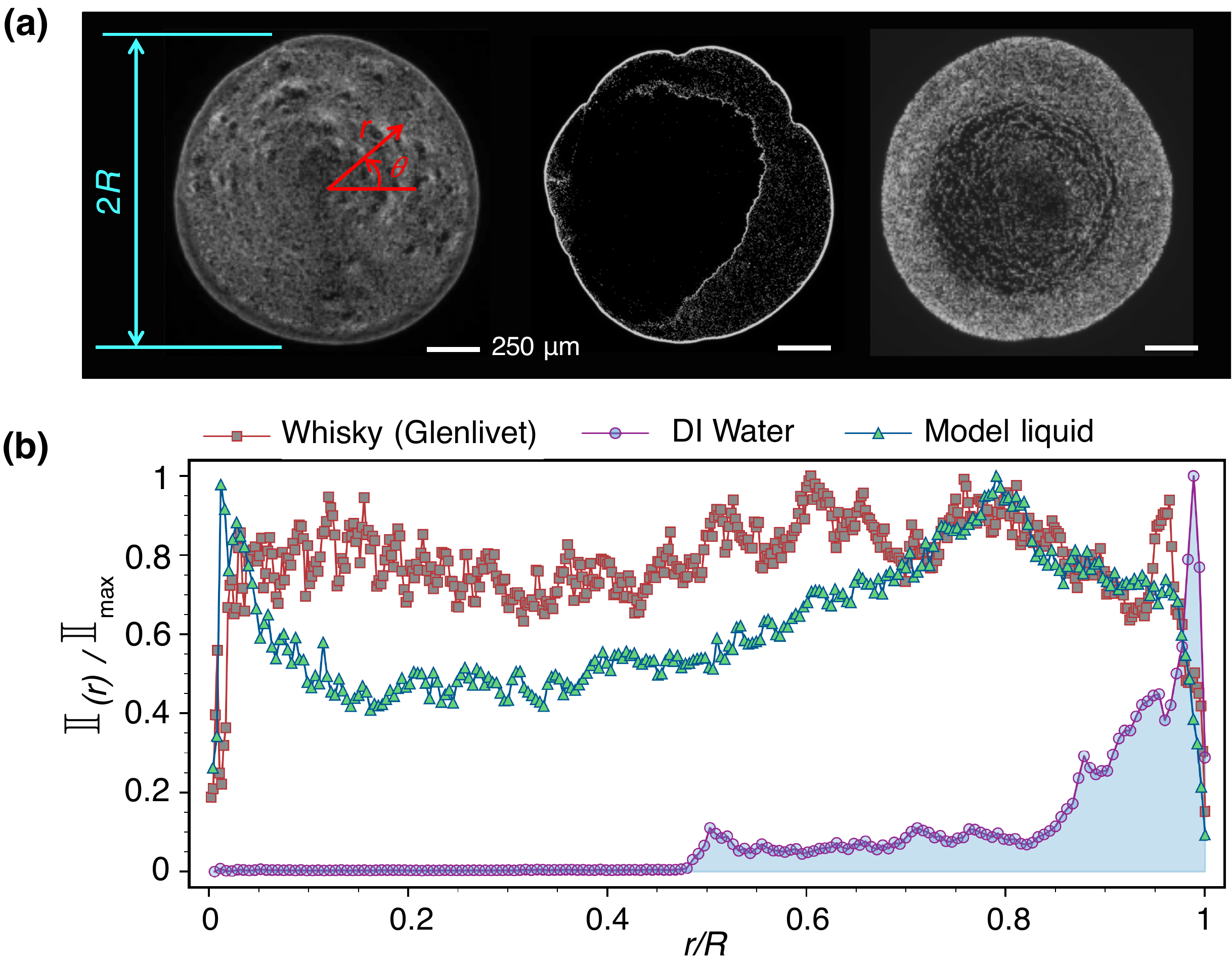}
  \caption{(a) Comparison of the final deposition patterns on top of the slide glass (Thermo Scientific, USA): (Left) whisky (Glenlivet, UK), (Center) water, and (Right) a model liquid (ethanol:water (35 wt\% : 35 wt\%) + SDS (0.05 wt\%) + PEO (0.05 wt\%)) containing carboxylate-modified fluorescent particles (diameter $d$ = 1 $\upmu$m). 2$R$ is the diameter of the final area. The particle concentration is 5 $\times$ 10$^{-3}$ vol\%. The scale bars represent 250 $\upmu$m. (b) Deposition profiles along $r$ are plotted for each of the images in (a). The intensity profile $\mathbb{I}(r)$ is normalized with the maximum of $\mathbb{I}(r)$ where $\mathbb{I}(r)$ = $\frac{1}{2\pi}$$\int_{0}^{2\pi} i(r,\theta) d\theta$ and $i(r,\theta)$ is the local light intensity.}
  \label{deposition2}
\end{figure*}

\end{document}